\documentclass[12pt]{article}
\usepackage{graphicx,color,overpic}
\usepackage{algpseudocode}
\usepackage{amsmath}
\usepackage{amsthm}
\usepackage{bm}
\usepackage{amssymb}
\usepackage{enumerate}
\usepackage{graphicx,overpic}
\usepackage{lscape}
\usepackage{listings}
\usepackage{color}
\usepackage{multirow}
\usepackage{booktabs}
\usepackage{setspace}
\usepackage{array}
\usepackage{listings}
\usepackage{float}
\usepackage{bm}
\usepackage[round]{natbib}
\usepackage{booktabs}
\usepackage[titletoc]{appendix}
\usepackage{multirow}
\usepackage{subfigure}
\usepackage{mathrsfs}
\usepackage{graphicx}
\usepackage{appendix}

\newtheorem{theorem}{Theorem}

\newcommand{\ba}{\begin{array}}
\newcommand{\ea}{\end{array}}
\newcommand{\bt}{\begin{tabular}}
\newcommand{\et}{\end{tabular}}
\newcommand{\btb}{\begin{table}}
\newcommand{\etb}{\end{table}}
\newcommand{\bc}{\begin{center}}
\newcommand{\ec}{\end{center}}
\newcommand{\bea}{\begin{eqnarray}}
\newcommand{\eea}{\end{eqnarray}}
\newcommand{\Bea}{\begin{eqnarray*}}
\newcommand{\Eea}{\end{eqnarray*}}
\newcommand{\beq}{\begin{equation}}
\newcommand{\eeq}{\end{equation}}

 \def \s2{{\sigma^2}}

\allowdisplaybreaks[4]
\begin{document}
\bibliographystyle{apalike}
\baselineskip 17pt
\title{\Large\bf Estimating the reciprocal of a binomial proportion}

{\small
\author{{\normalsize Jiajin Wei$^{1}, $ Ping He$^{2,} $\thanks{Co-corresponding author. E-mail: heping@uic.edu.cn}
~and Tiejun Tong$^{1, }$}\thanks{Co-corresponding author. E-mail: tongt@hkbu.edu.hk} \\\\
{\small $^1$Department of Mathematics, Hong Kong Baptist University, Hong Kong}\\
{\small $^2$Division of Science and Technology, BNU-HKBU United International College,}\\
{\small Zhuhai, China}\\
}
}

\maketitle \baselineskip 16pt

\begin{abstract} 
\noindent The binomial proportion is a classic parameter with many applications and has also been extensively studied in the literature. On the contrary, the reciprocal of the binomial proportion, or the inverse proportion, is often overlooked, even though it also
plays an important role in various fields including  clinical studies and random sampling. To estimate the inverse proportion, the maximum likelihood method, however, suffers from the zero-event problem. To overcome it, alternative methods have also been developed in the literature. Nevertheless, there is little work on investigating the theoretical behavior of the alternative methods, as well as on assessing their practical performance. In this paper, we first review the existing estimators for the inverse proportion and study their statistical properties, and then develop an optimal estimator within a family of shrinkage estimators. 
By simulation studies, we evaluate the performance of the existing and new estimators that aims to provide  the best estimators for practical use. 
We also revisit a recent meta-analysis on COVID-19 data to assess the relative risks of physical distancing on the infection of coronavirus, in which six out of seven studies encounter the zero-event problem.
\\\\
\noindent
$Key \ words$:
Binomial proportion,
Inverse proportion,
Relative risk,
Shrinkage estimator,
Zero-event problem

\end{abstract}

\newpage

\begin{spacing}{2}

\section{Introduction}\label{sec-1}
The binomial distribution is one of the most important distributions in statistics, which has been extensively studied in the literature with a wide range of applications.
 This classical distribution has two parameters $n$ and $p$, where $n$ is the number of independent Bernoulli trials and $p$ is the probability of success in each trial \citep{H05}. The probability of success, $p$, is also referred to as the binomial proportion. For excellent reviews on its estimation and inference, one may refer to, for example, \cite{AC98} and \cite{BCD01}.

Apart from the parameter $p$, it is known that some of its functions, say $p(1-p)$ and {\rm ln}$(p)$, also play important roles in statistics and have received much attention. In this article, we are interested in the reciprocal function
\begin{eqnarray}
\theta=\frac 1 p,
\label{inversep}
\end{eqnarray}
which is another important function of $p$ yet is often overlooked in the literature. For convenience, we also refer to $\theta$ in formula (\ref{inversep}) as the inverse proportion of the binomial distribution.
To demonstrate its usefulness, we will introduce some motivating examples in Section \ref{sec-F} that connect the inverse proportion with the relative risk (RR) and with the Horvitz-Thompson estimator \citep{HT52, F06}. Moreover, we will also introduce in Section \ref{dis} a relationship of the inverse proportion to the number needed to treat (NNT)
and the reduction in number to treat (RNT) in clinical studies, and present some future directions \citep{LSR88, A98, H00, Zhang2021}.

To start with,  let $X=\sum_{i=1}^n X_i$, where $X_i$ are independent and identically distributed  random variables from a Bernoulli distribution with success probability $p \in (0,1)$.
Then equivalently, $X$ follows a binomial distribution with parameters $n \geq 1$ and $p$.
Now if we want to estimate the inverse proportion $\theta$, a simple method will be to apply the maximum likelihood estimation (MLE) and it yields
\begin{eqnarray}
 \hat{\theta}_{{\tiny {\rm MLE}}}=\frac n X.
 \label{thetamle}
\end{eqnarray}
 This estimator is, however, not a valid estimator because it is not defined when $X=0$, i.e. when there is no successful event in $n$ trials. We refer to this problem as \textit{the zero-event problem} in the point estimation of $\theta$. In fact, the same problem also exists in the interval estimation of $p$. Specifically by
\cite{H05},
the $100(1-\alpha)$\% Wald interval is given as
\begin{eqnarray*}
\hat p \pm z_{\alpha/2}\sqrt{\frac{\hat p(1-\hat p)}{n}},
\end{eqnarray*}
where $\hat p=X/n$, and $z_{\alpha/2}$ is the upper $\alpha/2$ percentile of the standard normal distribution. When $X=0$, the lower and upper limits of the Wald interval are both zero; and consequently, they will not be able to provide a $(1-\alpha)$ coverage probability for the true proportion.

To overcome the zero-event problem, \cite{HL83} proposed the ``Rule of Three" to approximate the upper limit of the $95\%$
confidence interval (CI) for $p$.
Specifically, since the upper limit of the one-sided CI for $p$ is $1-0.05^{1/n}$ when $X=0$, the authors suggested to approximate this upper limit by $3/n$,
which then yields the simplified CI as $(0,3/n)$.
For more discussion on the ``Rule of Three", one may refer to \cite{TGM09} and the references therein. In particular, we note that the Wilson interval \citep{W27} and the Agresti-Coull interval \citep{AC98} for $p$ have also been referred to as the variations of the ``Rule of Three".

The Wilson interval was originated from Laplace who proposed the ``Law of Succession" in the 18th century. As mentioned in \cite{Good80}, Laplace's estimator for the binomial proportion was given as $(X+1)/(n+2)$, which is indeed a shrinkage estimator for $p$.
\cite{W27} generalized the shrinkage idea and proposed an updated ``Law of Succession" as $\tilde p(c)=(X+c)/(n+2c)$, where $c>0$ is a regularization parameter. Following the Wilson estimator, \cite{AC98} proposed to substitute $\tilde p(c)$ for $\hat p$ in the Wald interval  and yields the Agresti-Coull interval as
\begin{eqnarray*}
\tilde p(c) \pm z_{\alpha/2}\sqrt{\frac{\tilde p(c)[1-\tilde p(c)]}{n}}.
\end{eqnarray*}
It is also noteworthy that the Agresti-Coull interval always performs better than the Wald interval, no matter whether $n$ is large or small  \citep{BCD01}.

By applying the Wilson estimator $\tilde p$, one may estimate the inverse proportion as
\begin{eqnarray}
\tilde{\theta}_{}(c)=\frac{n+2c}{X+c}, \quad \quad c>0.
\label{est1}
\end{eqnarray}
Note that  the estimator with form (\ref{est1}) does not suffer from the zero-event problem, and so  provides a valid estimate of $\theta$ for any given $c>0$. In particular, two special cases of estimator (\ref{est1}) with $c=0.5$ and 1 have been widely applied  in the previous literature \citep{Walter1975, Carter2010}. Moreover, 
there are other estimators that follow the structure of (\ref{est1}) including, for example, a piecewise estimator (PE) with corrections only on $X=0$ or $n$ \citep{Schwarzer2007}.
In addition to  (\ref{est1}), another family of shrinkage estimators for the inverse proportion takes the form of 
\begin{eqnarray}
\hat\theta(c)=\frac{n+c}{X+c}, \quad \quad c>0.
\label{theta_c}
\end{eqnarray}
For the special case  $\hat\theta(0.5)$, it has  been  investigated by \cite{Pettigrew1986} and \cite{Hartung2001}.
More recently, \cite{F06} applied $\hat\theta(1)$ to estimate $\theta$ in sampling designs and demonstrated that it provides a good performance when $n$ is large. More specifically, it can be shown that $\hat\theta(1)$ is an asymptotically unbiased estimator of $\theta$ as $n$ tends to infinity \citep{chao1972negative, S13}.

In this paper, we first review the existing estimators for the inverse proportion and study their statistical properties, and then develop an optimal estimator within family 
(\ref{theta_c}).
In Section 2,  we briefly review the literature and introduce two real situations where an estimate of the inverse proportion is needed.
In Section 3, we derive the asymptotic properties of the existing estimators and derive the optimal shrinkage estimator within family (\ref{theta_c}). In Section \ref{4}, we conduct simulation studies to evaluate the performance of our new estimator, and compare it with existing competitors. In Section \ref{realdata}, we revisit a recent meta-analysis on COVID-19 data by \cite{chu2020} for assessing the relative risks of physical distancing on the infection of coronavirus, and then apply our new estimator to overcome the zero-event problem on the relative risks.
Lastly, we conclude the paper in Section 6 with some discussion and future work, and postpone the technical results in the Appendix.

\section{Literature review}\label{sec-F}

In this section, we provide two motivating examples in which an accurate  estimate of the inverse proportion $\theta$ is highly desired. 

\subsection{The relative risk}\label{rror}

In clinical studies, the relative risk (RR),  also known as the risk ratio, is a commonly used  effect size for measuring the effectiveness of a treatment or intervention.
Specifically,
RR is defined as
\begin{eqnarray}
{\rm RR} = \frac{p_1}{p_2},
\label{rrformula}
\end{eqnarray}
where $p_1$ is the event probability in the exposed group, and $p_2$ is the event probability in the unexposed group.

To estimate RR, we assume that there are $n_1$ samples in the exposed group with $X_1$ being the number of events, and $n_2$ samples in the unexposed group with $X_2$ being the number of events. Let also  $X_1$ follow a binomial distribution with parameters $n_1$ and $p_1$, $X_2$ follow a binomial distribution with parameters $n_2$ and $p_2$, and that they are independent of each other. Then by (\ref{rrformula}) and applying the MLEs of $p_1$ and $p_2$ respectively, RR can be estimated by
\begin{eqnarray}
\widehat{\rm RR} = {X_1/n_1 \over X_2/n_2} =  \frac{X_1n_2}{X_2n_1}.
\label{RRmle}
\end{eqnarray}
A problem of this estimator is, however, that it  suffers from the zero-event problem when $X_2=0$, which is the same problem as mentioned in Section 1 \citep{Wei2021}. To overcome this problem, there are a few popular suggestions in the literature to further improve the RR estimator in (\ref{RRmle}).  
\begin{enumerate}
\item[(\romannumeral1)] 
\cite{Walter1975} introduced a modified estimator of RR as $\widetilde{\rm RR}(0.5) = (X_1+0.5)(n_2+1)/[(X_2+0.5)(n_1+1)]$.
Following this idea, the inverse proportion of the unexposed group is, in fact, estimated by the Walter estimator
\begin{eqnarray}\label{Walter}
\tilde\theta(0.5) = \dfrac{n_2+1}{X_2+0.5}, 
\end{eqnarray}
which is a special case of estimator (\ref{est1}) with $c=0.5$.
\item[(\romannumeral2)] 
\cite{Pettigrew1986} proposed to estimate $p_i$ by $(X_i+0.5)/(n_i+0.5)$ for $i=1$ or $2$, and further concluded that 
${\rm ln}[(X_i+0.5)/(n_i+0.5)]$ is an unbiased estimator of ${\rm ln}(p_i)$ by ignoring the term $O(n^{-2})$. Accordingly, the Pettigrew estimator for the inverse proportion can be given as
\begin{eqnarray}\label{Pettigrew}
\hat\theta(0.5) = \dfrac{n_2+0.5}{X_2+0.5}, 
\end{eqnarray}
which is a special case of estimator (\ref{theta_c}) with $c=0.5$.

\item[(\romannumeral3)] 
Originated from (\ref{est1}), a family of piecewise estimators is defined as
\begin{eqnarray} \label{PEc}
  \tilde\theta_{\rm PE}(c) = \dfrac{n+2c I(X=0~{\rm or}~n)}{X+c I(X=0~{\rm or}~n)}, \quad \quad c>0,
\end{eqnarray}
where $I(\cdot)$ is the indicator function.  Particularly, one special case with $c=0.5$ that has been extensively applied in clinical studies \citep{Carter2010, Higgins2019, chu2020} is given as
\begin{eqnarray}\label{PiecewiseWalter}
\tilde\theta_{\rm PE}(0.5) = \dfrac{n_2+ I(X=0~{\rm or}~n)}{X_2+0.5 I(X=0~{\rm or}~n)}.
\end{eqnarray}
For ease of notation, we correspondingly denote this estimator as the piecewise Walter estimator in this paper.

\item[(\romannumeral4)] 
To further advance the piecewise Walter estimator,  \cite{Carter2010} proposed $\widetilde{\rm RR}(1) = (X_1+1)(n_2+2)/[(X_2+1)(n_1+2)]$, and it yields the Carter estimator for the inverse proportion
\begin{eqnarray}\label{Carter}
\tilde\theta(1) = \dfrac{n_2+2}{X_2+1},
\end{eqnarray}
which is a special case of estimator (\ref{est1}) with $c=1$.

\end{enumerate}

\subsection{The Horvitz-Thompson estimator}\label{sec-htestimator}

On random sampling without replacement from a finite population, it is known that the Horvitz-Thompson estimator has played  an important role in the literature for estimating the population total \citep{HT52, cochran2007}.

Let $U$ be a population composed of $t$ units $\{u_1, \dots, u_t\}$, and $p_i$ be the first-order selection probability associated with unit $u_i$. Let also $\Omega$ be  a random variable associated with the population $U$, and $\Omega_i$ be the value of $\Omega$ determined by unit $u_i$. Following these notations, the population total of $\Omega$ can be defined as $T=\sum_{i=1}^t \Omega_i$.
Then as an unbiased estimator of $T$, the Horvitz-Thompson estimator is given as
\begin{eqnarray}
\hat{T}=\sum_{j \in V} \omega_j \theta_j =\sum_{j \in V} \frac{\omega_j}{p_j},
\label{aaa2}
\end{eqnarray}
where $\omega_j$ is the observed value of $\Omega_j$, and $V\subseteq \{1,\dots,t\}$ is a subset of samples selected for estimating the population total. In practice, the inverse proportions $\theta_j=1/p_j$  are often unknown and need to be estimated.

To estimate $\theta_j$ in (\ref{aaa2}), \cite{F06} proposed a numerical method via Monte Carlo simulations. Specifically in each simulation, a total of $n$ samples were selected independently with replacement from the population $U$, with $X_j$ being the number of samples that contain the $j$th unit, where $j \in V$.
Further to avoid the zero-event problem on $X_j$, Fattorini applied estimator (\ref{theta_c}) with $c=1$ to estimate the inverse proportions by
\begin{eqnarray}
\hat\theta_{j}(1)=\frac{n+1}{X_j+1}, ~~~~~j\in V,
\label{theta1}
\end{eqnarray}
which then yields the modified Horvitz-Thompson estimator $\hat{T}_m$ as
$\hat{T}_m=\sum_{j \in V} \omega_j \hat \theta_j(1)$.
Unless otherwise specified, we will ignore the subscript $j$ in (\ref{theta1}) and refer to $\hat\theta(1)$ as the Fattorini estimator.


For the Fattorini estimator in family (\ref{theta_c}) with $c=1$, Seber (2013) showed that
 \begin{eqnarray}
{E}[\hat\theta_{}(1)] ={E}\left(\frac{n+1}{X+1} \right)=\frac{1-(1-p)^{n+1}}{p}=\theta-\theta\left(1-{1\over \theta}\right)^{n+1}.
 \label{E_Fattorini}
 \end{eqnarray}
Then by the fact that $\lim_{n \to \infty} {\rm Bias}_{}[\hat\theta_{}(1)]=\lim_{n \to \infty} [-\theta(1-1/\theta)^{n+1}] = 0$  for any fixed $\theta\in (1,\infty)$, the Fattorini estimator is an asymptotically unbiased estimator of $\theta$ when $n$ is large.
In addition, when $p$ is large enough, or equivalently when $\theta$ is close to 1, the estimation bias of the Fattorini estimator is often negligible no matter whether $n$ is large or small.

\section{Methodology}\label{sec-4}

\subsection{Comparison of the existing estimators} \label{sec-3-1}

In view of the demand for accurate estimation of the inverse proportion, we revisit the three families of shrinkage estimators in (\ref{est1}), (\ref{theta_c}) and (\ref{PEc})  and  compare them in both theory and practice. We first show that the three estimators are all consistent and asymptotically equivalent, with the proof of the theorem in Appendix A. 
\vspace{0.3cm}
\begin{theorem}\label{theorem1}
Let $X$ be a binomial random variable with parameters $n$ and $p$.
For the shrinkage estimators in (\ref{est1}), (\ref{theta_c}) and (\ref{PEc}) with any finite $c>0$, we have the following properties: 
\begin{enumerate}
\item[(\romannumeral1)] $\tilde{\theta}_{}(c)$, $\hat\theta(c)$ and $\tilde\theta_{\rm PE}(c)$ are all consistent estimators of $\theta$; 
\item[(\romannumeral2)]  $\tilde{\theta}_{}(c)$, $\hat\theta(c)$ and $\tilde\theta_{\rm PE}(c)$ are all asymptotically equivalent such that $\sqrt{n}(\check\theta- \theta)  \stackrel{D}{\rightarrow} N(0, \theta^2(\theta-1))$, where $\check\theta$ is a generic notation for the three estimators and $\stackrel{D}{\rightarrow}$ denotes convergence in distribution.  
\end{enumerate}
\end{theorem}
\vspace{0.3cm}

\begin{figure}[!hbt]
\centering
\includegraphics[width=15.1cm]{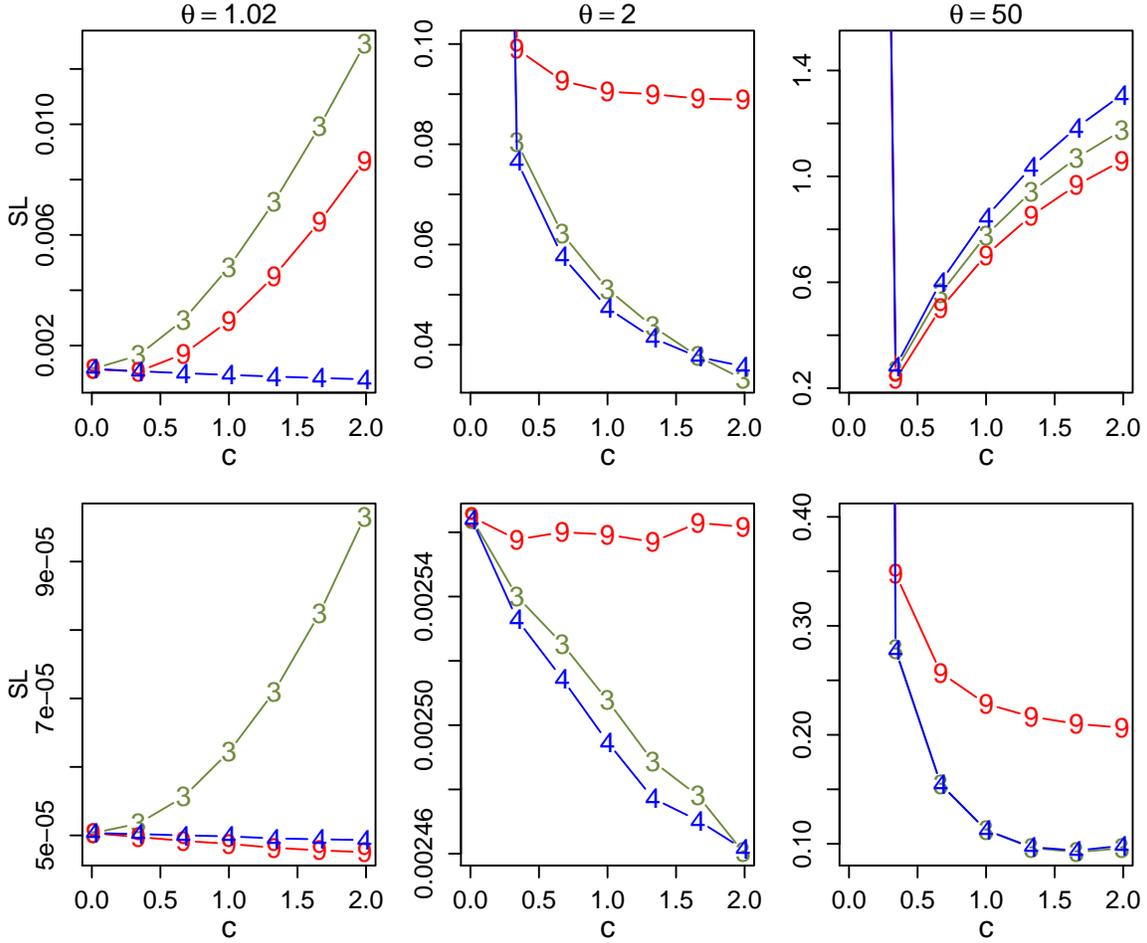}
\caption{The Stein losses for the shrinkage estimators from the three families  with $\theta=$ 1.02, 2 or 50, $n=$ 10 (top three panels) or 200 (bottom three panels), and $c\in (0,2)$, where 
``3" represents the estimators from family (\ref{est1}),
``4" represents the estimators from family (\ref{theta_c}),
and ``9" represents the estimators from family (\ref{PEc}).
}
\label{Figure-1}
\end{figure}

Despite the asymptotic equivalence, we note however that their finite-sample performance can be quite different. To illustrate it, we conduct a numerical study by considering $\theta=$ 1.02, 2 or 50, which is equivalent to $p=$ 0.98, 0.5 or 0.02. We also consider $n=$ 10 or 200 to represent the small and large sample sizes respectively, and let $c$ range from 0 to 2 so as to cover most common choices of $c$ in the literature. Then for each setting, we generate $N=1,000,000$ data sets from the binomial distribution and estimate $\theta$ by each estimator  from the three families. Finally, with the simulated data sets, we compute the Stein loss (SL) \citep{Dey1985} of each estimator by 
\begin{eqnarray}\label{SL}
{\rm SL}(\check \theta_k) = \frac{1}{N} \sum_{k=1}^N \left[ \frac{\check \theta_k}{\theta} - {\rm ln} \left(\frac{\check \theta_k}{\theta} \right) -1 \right],
\end{eqnarray}
and then report the simulation results in Figure \ref{Figure-1}. 

From Figure \ref{Figure-1}, it is evident that the estimators from family (\ref{theta_c}) perform better than those from the other two families in most settings.
In particular, no estimator from family (\ref{est1}) is able to provide an accurate estimate when $\theta=1.02$, no matter whether the sample size  is large or small.
On the other side, the estimators from family (\ref{PEc}) fail to provide a stable performance when $\theta$ is moderate to large. To summarize, except for the extreme case where $\theta$ is relatively large and $n$ is relatively small, the estimators from family (\ref{theta_c})  are always among the best and so can be safely recommended. 
Moreover, we also provide a theoretical evidence from the perspective of bias that the estimators from family (\ref{est1}) can be suboptimal for practical use.

\vspace{0.3cm}
\begin{theorem}\label{theorem2} 
Let $X$ be a binomial random variable with parameters $n $ and $p$. Then for the estimators from family  (\ref{est1}), there does not exist a shrinkage parameter $c$ such that $E[\tilde{\theta}_{}(c)]=\theta$ when $p=0.5$, or equivalently, when $\theta=2$.
\end{theorem}
\vspace{0.3cm}

The proof of Theorem 2 is given in Appendix B. Taken together the above comparisons, we propose to probe into the family of estimators (\ref{theta_c}) and find the optimal estimator of $\theta$ in this paper.

\subsection{Optimal estimation of $\bm{\theta}$} \label{sec-4.1}


For the estimators from family (\ref{theta_c}), we have introduced the Fattorini estimator with $c=1$ as a special case with the asymptotic property  in Section \ref{sec-htestimator}. However, as is shown in the numerical study, the Fattorini estimator may not provide an accurate estimate for the inverse proportion when $n$ is small and $\theta$ is large.
To further illustrate it, we take $n=10$ and $\theta=50$; then according to (\ref{E_Fattorini}), the relative bias of the Fattorini estimator is as large as
\begin{eqnarray*}
\frac{E[\hat\theta(1)] - \theta}{\theta} \times 100\%
=-(1-0.02)^{11} \times 100\% \approx -80.07 \%.
\end{eqnarray*}
 In addition, it is noteworthy that the expected value of the Fattorini estimator is always lower than $\theta$ and so is consistently negatively biased.
These evidences indicate that the Fattorini estimator may not be the optimal estimator in family (\ref{theta_c}).
 
To alleviate the bias in the Fattorini estimator, we now define the optimal shrinkage parameter $c$ as the value  such that  ${E}[\hat\theta(c)] = \theta$.
For ease of notation, we also express the expected value of $\hat\theta(c)$ as
\begin{eqnarray}
g(c)={E}[\hat\theta(c)]
=\sum_{x=0}^n \left(\frac{n+c}{x+c}\right) {n\choose x} p^x (1-p)^{n-x},
\label{g_c}
\end{eqnarray}
and then regard $g(c)$ as a function of $c$. In the following theorem, we provide some properties of $g(c)$, including the continuity, monotonicity and convexity, with the proof in Appendix C.  
\vspace{0.3cm}
\begin{theorem}\label{theorem3}
For the expected value function $g(c)$ in (\ref{g_c}) with any finite integer $n$, we have the following properties:
\begin{enumerate}
\item[(\romannumeral1)]  $g(c)$ is a continuous function of $c$ on $(0, \infty)$ with $\lim_{c\to 0} g(c)=\infty$ \\and $\lim_{c\to \infty} g(c)=1$;
\item[(\romannumeral2)]  $g(c)$ is a strictly decreasing function of $c$ on $(0, \infty)$;
\item[(\romannumeral3)]  $g(c)$ is a strictly convex function of $c$ on $(0, \infty)$.
\end{enumerate}
\end{theorem}
\vspace{0.3cm}

Note also that $\theta$ takes value on $(1,\infty)$, and $g(1)<\theta$ for any fixed $n$ according to  formula (\ref{E_Fattorini}). Then by Theorem \ref{theorem3} and the  Intermediate Value Theorem, there exists a unique solution $c \in (0,1)$ such that $g(c)=\theta$, or equivalently,
\begin{align}
g(c)=
\sum_{x=0}^n \left(\frac{n+c}{x+c}\right) {n\choose x} p^x (1-p)^{n-x}=\frac{1}{p}.
\label{a7}
\end{align}
When $n$ is small,  in particular for $n=1$ or $n=2$, we can derive the explicit solution of $c$ from equation (\ref{a7}). When $n$ is large, since the degree of equation as a function of $c$ is with $n+1$, there may not have an explicit solution for $c$ in mathematics.
To summarize, we have the following theorem with the proof in Appendix D.
\vspace{0.3cm}
\begin{theorem}\label{theorem4} 
When $n$ is less than 3, the solution of $c$ in equation (\ref{a7}) is given by
\begin{eqnarray*}
c_n = \left\{
\begin{array}{ll}
p          & ~~~~~~~~ n=1,  \\
p- {0.5} + \sqrt{{0.5} - (p-{0.5})^2} & ~~~~~~~~ n=2.
\end{array}
\right.
\end{eqnarray*}
When $n\geq3$, we have the approximate solution of $c$ as
\begin{eqnarray}
c_n \approx 1 - {p^{-1}(1-p)^{n+1} \over  (n+1)(1+D_1)D_2-D_1},
\label{theorem3_formula}
\end{eqnarray}
where
\begin{eqnarray*}
&&D_1 = {1\over p(n+1)}[1-(1-p)^{n+1}], \\
&&D_2 = {1\over p^2(n+1)(n+2)}[1-(1-p)^{n+2}-(n+2)p(1-p)^{n+1}].
\end{eqnarray*}
\end{theorem}
\vspace{0.3cm}

\begin{figure}[!hbt]
\centering
\includegraphics[width=12cm]{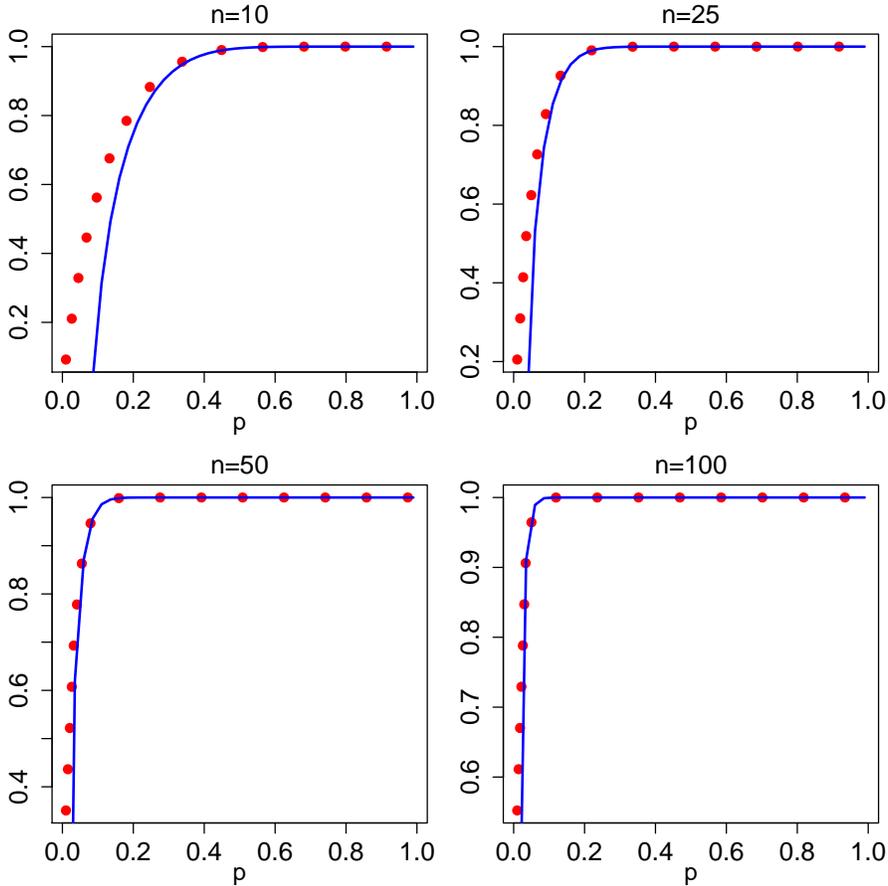}
\caption{The true and approximate solutions of $c$ with $n=$10, 25, 50 or 100. The solid dots represent the values of the true solution, and the solid lines represent the values of the approximate solution.}
\label{Figure_approx_c}
\end{figure}

To check the accuracy of the approximate solution in Theorem \ref{theorem4}, we also plot  the numerical results of the true and approximate solutions of $c$ as a function of $p$ in Figure \ref{Figure_approx_c}. Under various settings, we note that the true solution of $c$ is given as a monotonically increasing function of $p$ with the upper bound 1. And in addition, our approximate solution always works well as long as $n$ or $p$ is not extremely small.

\subsection{Plug-in estimator}\label{3.2}
To apply Theorem \ref{theorem4} for the optimal shrinkage parameter, we need a plug-in estimator for the unknown $p$. Intuitively, the MLE of $p$, $\hat p_{\text {\tiny {MLE}}}=X/n$, can serve as a natural choice. By doing so, however, for $n=1$ we have $\hat c_1=\hat p_{\text {\tiny {MLE}}} = X $ and further it yields that $\hat \theta(\hat c_1) = {(1+\hat c_1)}/{(X+\hat c_1)} = (1+X)/2X$, which then suffers from the zero-event problem. For $n=2$, it is noted that the same problem also remains.  While for $n\geq 3$, the approximate solution will no longer suffer from the zero-event problem; but on the other side, the denominator term, $(n+1)(1+D_1)D_2-D_1$, in (\ref{theorem3_formula}) will be zero when $X=n$, and consequently the approximate solution is still not be applicable. To conclude, the MLE of $p$ cannot be directly applied as  the plug-in estimator when applying Theorem \ref{theorem4} to estimate the inverse proportion.

\begin{figure}[!hbt]
\centering
\includegraphics[width=12cm]{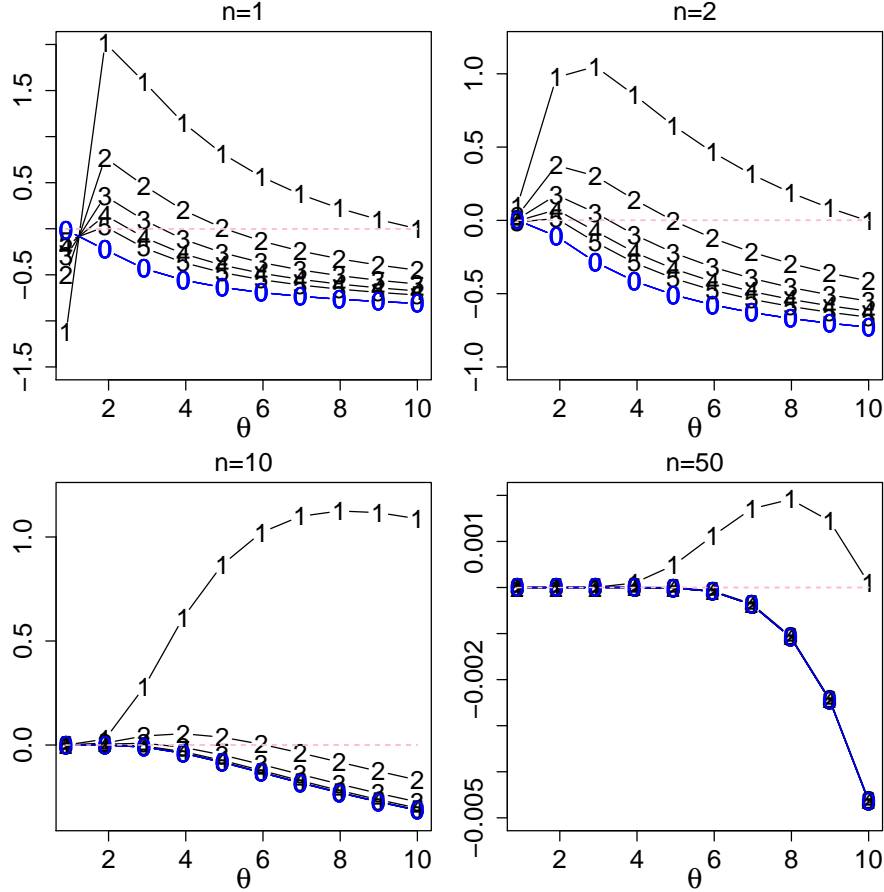} 
\caption{The relative biases of $\hat\theta_{}(\tilde c_n)$ with $\alpha=$ 0.1, 0.2, 0.3, 0.4 or 0.5, where ``1" represents the relative biases associated with $\alpha=$ 0.1, ``2" represents the relative biases associated with $\alpha=$ 0.2, ``3"  represents the relative biases associated with $\alpha=$ 0.3, ``4" represents the relative biases associated with $\alpha=$ 0.4, and ``5" represents the relative biases associated with $\alpha=$ 0.5. And for comparison, ``0" represents the relative biases of the Fattorini estimator.}
\label{alpha_compare}
\end{figure}

To overcome the boundary problems on both sides, we consider the plug-in estimator of $p$ with the following structure:
\begin{equation*}
\tilde p_{\text {plug}}(\alpha)={\rm min}({\rm max}(\hat p_{\text {\tiny {MLE}}}, \alpha), 1-\alpha),
\end{equation*}
where $0 < \alpha \leq 0.5$ is the threshold parameter. Then with $\tilde p_{\text {plug}}(\alpha)$ as the plug-in estimator of $p$, we let  $\tilde c_n(\alpha)$ be the estimator of $c_n$ in Theorem \ref{theorem4}.
To determine the best threshold value for practical use, we take several different  $\alpha$ and then compute the relative bias of the estimator by
\begin{eqnarray}\label{relativebias}
{\rm Bias}(\check\theta_k) =  \frac{1}{N}\sum_{k=1}^N \left({\check\theta_k \over \theta }-1\right),
\end{eqnarray}
 where $\check\theta_k$ is a generic form of $\hat\theta_{k}(\tilde c_n(\alpha))$.
Specifically in Figure 2, we plot the relative biases of the estimator as functions of $\theta$ for $\alpha=0.1$, 0.2, 0.3, 0.4, 0.5 and $n=1$, 2, 10, 50. While for comparison, the relative biases of the Fattorini estimator are also presented in Figure 2. 

In the top two panels of  Figure \ref{alpha_compare}, it is evident that a small threshold value, say $\alpha=0.1$ or 0.2, may not provide an adequate remedy for the boundary problems when $n$ is extremely small. Note also that  $\tilde p_{\text {plug}}(\alpha)=0.5$ when $\alpha=0.5$. Then by Figure 1 that $c_n$ is always close to 1 when $p=0.5$,  the resulting estimator of $\theta$ with $\alpha=0.5$ will be nearly the same as the Fattorini estimator when $n$ is large. And for moderate sample sizes, say $n=10$ and $n=50$, the bottom two panels of Figure \ref{alpha_compare} show that the best value of $\alpha$ should be neither too small or too large.
Taken together, we recommend to apply $\alpha_n = 1/(2+{\rm ln}(n))$ as the adaptive threshold value, which follows a decreasing trend, say, for example, $\alpha_1=0.5$, $\alpha_{10}=0.23$, $\alpha_{100}=0.15$, and $\alpha_{1000}=0.11$.
Then with $\tilde p_{\text {plug}}(\alpha_n) = {\rm min}({\rm max}(\hat p_{\text {\tiny {MLE}}}, \alpha_n), 1-\alpha_n)$ as the plug-in estimator, our final estimator of the inverse proportion is given by
\begin{eqnarray}
\hat\theta_{}(\tilde c_n)=\frac{n+ \tilde c_n}{X+ \tilde c_n},
\label{estimator_hat_c}
\end{eqnarray}
where $\tilde c_n = c_n(\tilde p_{\rm plug}(\alpha_n ))$ is the estimator of $c_n$ given in Theorem \ref{theorem4}. In addition, we derive the asymptotic properties of estimator (\ref{estimator_hat_c}) in the following theorem with the proof in Appendix E. 
\vspace{0.3cm}
\begin{theorem}\label{theorem5}
Let $X$ be a binomial random variable with parameters $n $ and $p$.
For the estimator $\hat\theta_{}(\tilde c_n)$ in (\ref{estimator_hat_c}), we have $\tilde c_n = 1 + o_p(1)$ and $\hat\theta_{}(\tilde c_n)$ is a consistent estimator of $\theta$.
\end{theorem}
\vspace{0.3cm}

\section{Simulation studies}\label{4}
\begin{figure}[!hbt]
\centering
\includegraphics[width=15.1cm]{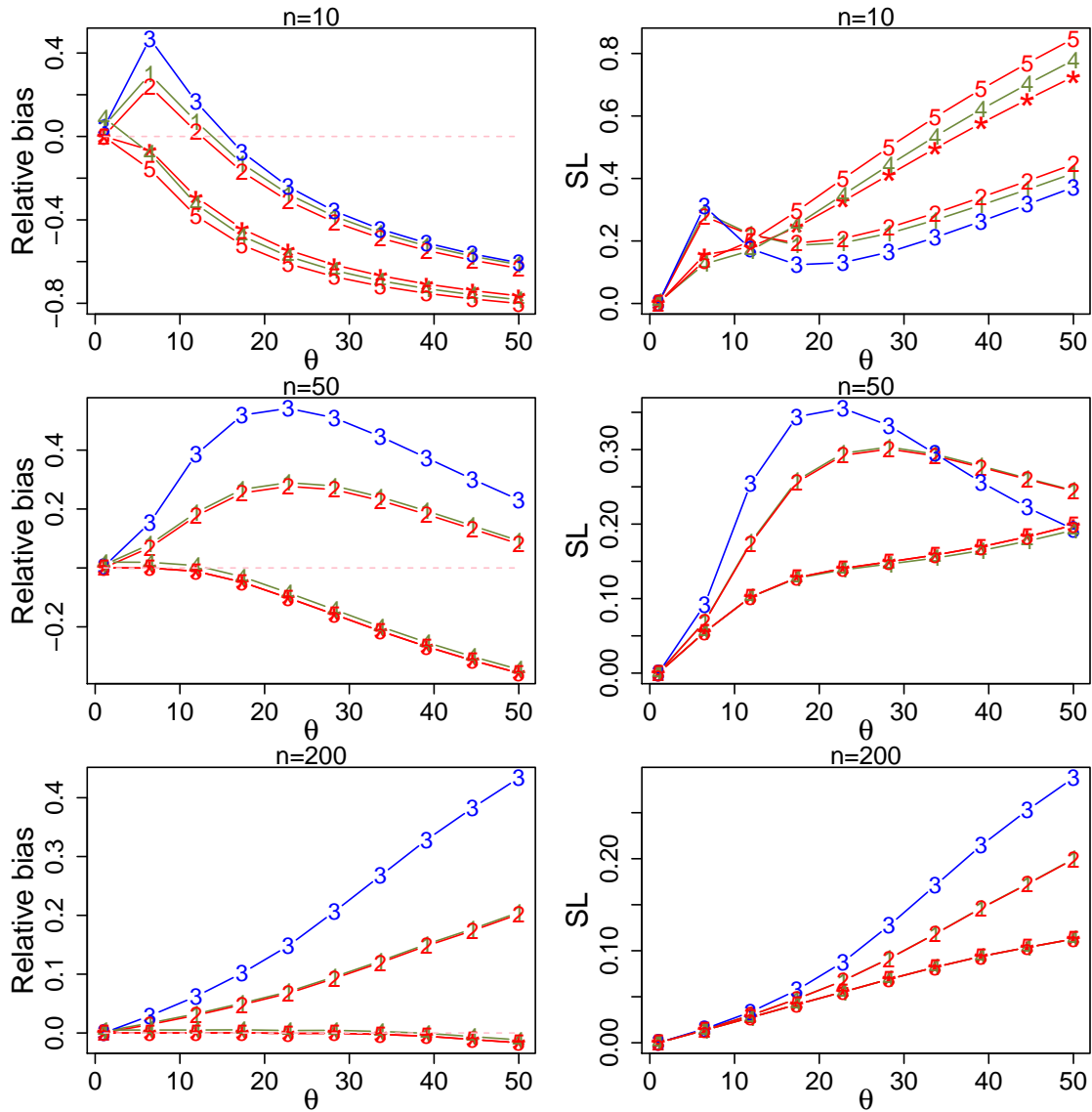}
 \caption{The relative biases and the Stein losses of the six estimators with $n=10$, 50 or 200, where ``\textasteriskcentered" represents our new estimator, ``1" represents the Walter estimator,
``2" represents the Pettigrew estimator, ``3" represents the piecewise Walter estimator, ``4" represents the Carter estimator, and ``5" represents the Fattorini estimator.}
\label{Figure-4}
\end{figure}

In this section, we conduct simulation studies to evaluate the finite-sample performance of our new estimator in (\ref{estimator_hat_c}) for the inverse proportion. 
For comparison, five existing estimators in the literature are also considered, including
the Walter estimator in (\ref{Walter}), the Pettigrew estimator in (\ref{Pettigrew}), the piecewise Walter estimator in (\ref{PiecewiseWalter}), 
the Carter estimator in (\ref{Carter}), and the Fattorini estimator in (\ref{theta1}). For the simulation settings, we let $\theta$ range from 1.02 up to 50, which is equivalent to $p$ ranging from 0.98 down to 0.02, and consider  $n=$ 10, 50 or 200 as three different sample sizes.
We further generate $N=1,000,000$ data sets from the binomial distribution with each combination of $\theta$ and $n$. Finally, we compute the relative bias by (\ref{relativebias}) and compute the Stein loss by (\ref{SL}) for each estimator, and  then report the simulation results in Figure \ref{Figure-4}. 

From Figure \ref{Figure-4}, it is evident that the new, Carter and Fattorini estimators perform comparably and yield reliable estimates in most settings. 
In contrast, the Walter, Pettigrew and piecewise Walter estimators fail to provide a stable performance, especially when  the sample size is large. 
When the sample size is small, say $n=10$, the new estimator outperforms the Carter and Fattorini estimators from the perspectives of both the relative biases and the Stein losses.
Moreover, from Appendix F, we note that the Carter estimator is alway the most biased estimator for small $\theta$ no matter whether the sample size is large or small. 
To conclude, the new estimator can serve as the most reliable estimator of the inverse proportion for practical use. In addition, if a slightly less accurate estimate is acceptable, then the Fattorini estimator can also be recommended by virtue of its simple form and the good performance when the sample size is reasonably large.

\section{An application to zero-event studies}\label{realdata}
In this section, we apply our new estimator into a  meta-analysis on COVID-19 data with zero-event studies.
\cite{chu2020} carried out an excellent review to investigate effects of physical distancing, face masks and eye protection on the infection of severe acute respiratory syndrome (SARS), Middle East respiratory syndrome (MERS) and severe acute respiratory syndrome coronavirus 2 (SARS-CoV-2).
This systematic review was published in June 2020 and is now attracting more and more attention, for example in Google Scholar as of 7 June 2021, their paper has already received a total of 1682	 citations.
Also as commented by \cite{MW20}, this systematic review provides a landmark for people to be aware of the importance of physical distancing and face protection.
In particular for physical distancing, they applied the relative risks as effect sizes and concluded that the virus transmission is significantly reduced with a further distance.

\begin{figure}[!hbt]
\centering
\includegraphics[width=15.5cm]{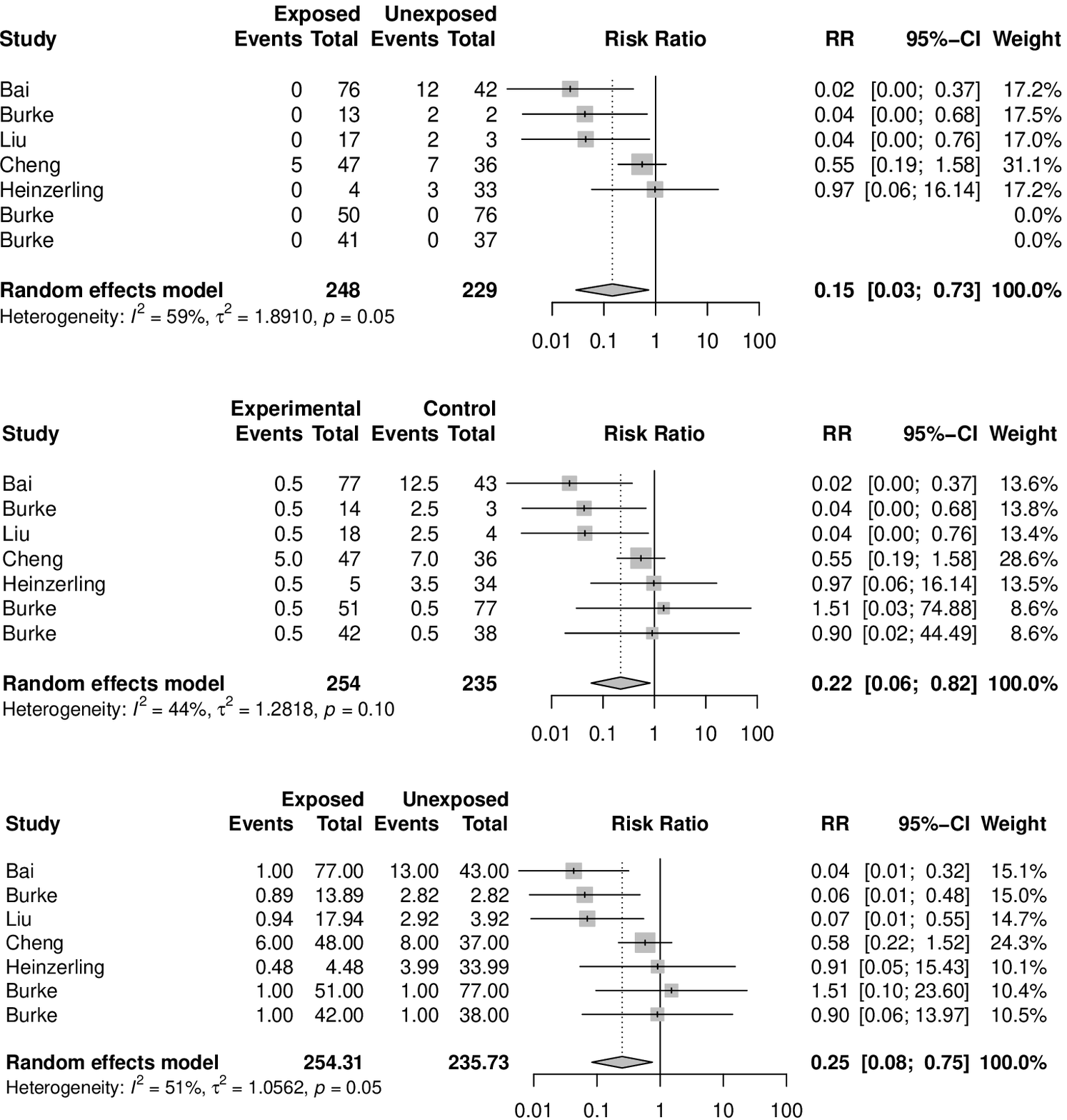}
\caption{Forest plots on the relative risk between physical distancing and infection for COVID-19 data.}
\label{Figure-5}
\end{figure}

In the top panel of Figure \ref{Figure-5}, seven studies were included in their meta-analysis of physical distancing for COVID-19 data, where six studies therein suffered from the zero-event problem.
For the four single-zero-event studies, the 0.5 continuity correction was added to all the counts of events, while for the two double-zero-event studies, they were not included in the meta-analysis.
By \cite{xu2020} and our simulation results, adding the 0.5 continuity correction is suboptimal. Moreover, \cite{xu2020}  also showed that the double-zero-event studies may also be  informative, and so excluding them can be questionable and/or even alter the results. In view of the above limitations, we re-conducted the meta-analysis on
COVID-19 data that also includes the two double-zero-event studies. Specifically, by applying our new estimator in (\ref{estimator_hat_c}),  the relative risks are estimated by
\begin{eqnarray}
\widehat{\rm RR}(\tilde c_n) = \frac{(X_1+\tilde c_{n_1})(n_2+\tilde c_{n_2})}{(X_2+\tilde c_{n_2})(n_1+\tilde c_{n_1})},
\label{newRRestimator}
\end{eqnarray}
where $\tilde c_{n_1}$ and $\tilde c_{n_2}$ are the estimates of the optimal shrinkage parameter for the exposed group and the unexposed group, respectively. While for comparison, we also conduct a meta-analysis for all seven studies by  the 0.5 continuity correction, and then present all the forest plots in Figure \ref{Figure-5}.

From the middle and bottom panels of Figure \ref{Figure-5}, it is evident that the new meta-analytical results with the double-zero-event studies also support the claim that a further distance will reduce the virus infection. On the other hand, the evidence becomes less significant as the combined relative risks get larger. Moreover, by comparing the two forest plots that both include the double-zero-event studies, we also note that our new estimator in (\ref{newRRestimator}) is able to yield a larger combined relative risk with a narrower confidence interval. By the variance function of ${\rm ln(\widehat{RR})}$, $1/X_1-1/n_1+1/X_2-1/n_2$, the 0.5 continuity correction may lead to a large estimate of the relative risk after the exponential transformation, especially when the zero-event problem occurs. Hence,  the confidence intervals  of the relative risks in the two double-zero-event studies are very wide, which can indicate that there may exist high uncertainty in the interval estimation.  In contrast, by applying our new estimator of the inverse proportion, the confidence intervals for the double-zero-event studies will be much narrower.

\section{Conclusion}\label{dis}
The binomial proportion is a classic parameter originated from the binomial distribution, which has been well studied in the literature because of its wide range of applications. In contrast, the reciprocal of the binomial proportion, also known as the inverse proportion, is often overlooked, although
it also plays an important role in various fields including clinical studies and random sampling. However, it is known that the MLE of the inverse proportion suffers from the zero-event problem. To overcome this problem, there have been a number of existing estimators in the literature for the inverse proportion.

To further advance the literature, we first introduced two motivating examples where an accurate estimate of the inverse proportion is desired. We then compared three shrinkage families of estimators and figured out the family with better statistical properties. Finally, we  proposed a new estimator of the inverse proportion by deriving the optimal shrinkage parameter $c$ in the family of estimators (\ref{theta_c}). To be more specific, we derived the explicit formula for the optimal $c$ in Theorem \ref{theorem4} for  $n=1$ or 2, and an approximate formula for the optimal  $c$ for $n\geq3$.  Further to estimate the unknown $p$ in the formula of the optimal shrinkage parameter,
a plug-in estimator was also introduced and that also overcame the boundary problem of $p$. 
Simulation studies showed that our new estimator performs better than, or as well as, the existing competitors in most practical settings, and it can thus be recommended to estimate the inverse proportion for practical application.
 Finally, we also applied  our new estimator to a recent meta-analysis on COVID-19 data with the zero-event problem, and it yielded more reliable results for the scientific question how physical
distancing can effectively prevent the infection of the new coronavirus.

To conclude the paper, we have made a good effort in finding  the optimal estimator for the inverse proportion related to the binomial distribution. According to \cite{G67}, there does not exist an unbiased estimator for the inverse proportion $\theta$.
To verify this result,
by the proof-by-contradiction we assume that $\hat{\theta}_{\rm u} = \eta(X)$ is an unbiased estimator of $\theta$. Then by definition,
${E}(\hat{\theta}_{\rm u})=\sum_{x=0}^{n} \eta(x) {n\choose x} p^{x} (1-p)^{n-{x}}=\theta.$
From the left-hand side, the expected value of $\hat{\theta}_{\rm u}$ is a polynomial of $p$ with degree $n$.
While for the right-hand side, by the Taylor expansion we have $\theta=1/p=\sum_{i=0}^{\infty} (1-p)^i$, which is a polynomial of $p$ with infinite degree. This shows that the unbiasedness cannot be held for any finite $n$. In view of this property, there is probably no uniformly best estimator for the inverse proportion. Although we have conducted some nice work in this paper, we believe that
more advanced research is still needed to further improve the estimation accuracy of the inverse proportion. For example, one may consider to develop
a better and more robust approximation for  the optimal shrinkage parameter when the  binomial proportion $p$ is extremely small. In addition, other families of shrinkage estimators can also be considered to see whether they can yield better estimators for the inverse proportion.

Last but not least, we note that our new estimation of the inverse proportion can have many other real applications. For instance, the spirit of  our new method may also be applied to estimate the number needed to treat (NNT), which is another important medical term and was first introduced by  \cite{LSR88}.  Specifically, NNT is defined as ${\rm NNT} = 1/(p_1-p_2)$,
where $p_1$ is the event probability in the exposed group and $p_2$ is the event probability in the unexposed group.
Noting also that $p_1-p_2$ is the absolute risk reduction (ARR), NNT can be explained as the average number of patients who are needed to be treated to obtain one more patient cured compared with a control in a clinical trial \citep{H00}.
Nevertheless, the estimation of NNT will be more challenging than the estimation of the inverse proportion, mainly because  the estimate of $p_1-p_2$ can be either positive or negative, in addition to the zero-event problem in the denominator.  More recently, \cite{V19} also referred to this situation as the statistically nonsignificant result, which may lead to an unexpected  calculation complication.

In addition to NNT, \cite{Zhang2021} proposed the reduction in number to treat (RNT) as a new measure of the treatment effect in randomized control trials. Specifically, let the two inverse proportions $\theta_1 = 1/p_1$ be the average number of patients who are needed to be treated to obtain one patient cured in the exposed group and 
$\theta_2 = 1/p_2$ be the average number of patients who are needed to be treated to obtain one patient cured in the unexposed group, then RNT is defined as 
${\rm RNT}  = \theta_2 - \theta_1 = 1/p_2 - 1/p_1$.
Also by (\ref{thetamle}), the MLE of RNT is given as 
$\widehat{\rm RNT}_{\rm MLE}  =  n_2/X_2 - n_1/X_1$,
which once again may not be applicable when the value of $X_1$ or $X_2$ is zero. 
Thus to study  the statistical inference of RNT, it also requires a valid estimate for each of the  inverse proportions that does not suffer from the zero-event problem.
We expect that our new work in this paper will shed light on new directions on the NNT and RNT estimation, which can be particularly useful in clinical trials and evidence-based medicine.

\bibliographystyle{apalike}

\newpage

\section*{Appendix A: Proof of Theorem \ref{theorem1}}

{\it Proof}. To prove (i), the inverse of $\tilde\theta(c)=(n+2c)/(X+c)$ in (\ref{est1}) is given as 
\begin{eqnarray} \label{app_1.1}
\frac{X+c}{n+2c} = \left(\frac{n}{n+2c} \right) \frac{X}{n} + \frac{c}{n+2c}
\end{eqnarray} 
For any $p \in(0,1)$, we note that $X/n$ converges to $p$ in probability as $n \to \infty$.  Thus for any fixed $c>0$, by Slutsky's Theorem, formula (\ref{app_1.1}) also converges to $p$ in probability as $n \to \infty$. This shows that  $\tilde\theta(c)$ converges to $\theta$ in probability as $n \to \infty$ for any $\theta \in (1,\infty)$, i.e., $\tilde\theta(c)$ is a consistent estimator of $\theta$. The proofs for the other two estimators are similar and so are omitted for the sake of brevity.

To prove (ii), let $Y$ be a Bernoulli random variable and let $L(\theta|Y) = (1/\theta)^Y(1-1/\theta)^{1-Y}$ be the likelihood function.
Then the Fisher information $I(\theta)$ is 
\begin{eqnarray*}
I(\theta) 
&=& -E \left\{\frac{\partial^2}{\partial \theta^2} [  -Y{\rm ln}\theta +  (1-Y){\rm ln}(1-1/\theta)]\;\middle|\; \theta   \right\}\\
&=& -E \left(\frac{Y}{\theta^2} + \frac{(1-Y)(1-2\theta)}{(\theta^2-\theta)^2}  \;\middle|\; \theta   \right) \\
&=& -\frac{1}{\theta^3} - \frac{(1-1/\theta)(1-2\theta)}{(\theta^2-\theta)^2}\\
&=& \frac{1}{\theta^2(\theta-1)}.
\end{eqnarray*} 
By the asymptotic normality of the MLE, we have
\begin{eqnarray*}
\sqrt{n}(\hat\theta_{\rm MLE}-\theta) \stackrel{D}{\rightarrow} N(0, \theta^2(\theta-1) ~~~~~ {\rm as}~n \to \infty. 
\end{eqnarray*}

Then for  $\tilde\theta(c)=(n+2c)/(X+c)$ in (\ref{est1}), we note that 
\begin{eqnarray*}
\sqrt{n}[\tilde\theta(c)- \theta] &=&     \sqrt{n}(\hat\theta_{\rm MLE}-\theta) + \sqrt{n}[\tilde\theta(c)- \hat\theta_{\rm MLE}] \\
&=&   \sqrt{n}(\hat\theta_{\rm MLE}-\theta) + \sqrt{n} \frac{(2X-n)c}{(X+c)X} \\
&=&  \sqrt{n}(\hat\theta_{\rm MLE}-\theta) + \sqrt{n} \left( 2-\frac{n}{X}\right) \left( \frac{c}{X+c}\right) \\
&=&  \sqrt{n}(\hat\theta_{\rm MLE}-\theta) +  o_p(1).
\end{eqnarray*}
Then by Slutsky’s Theorem, it yields that
$
\sqrt{n}[\tilde\theta(c)- \theta]  \stackrel{D}{\rightarrow} N(0, \theta^2(\theta-1))
$
as $n \to \infty$. The proofs for the other two estimators are similar, and so are omitted. Consequently, the three estimators in (\ref{est1}), (\ref{theta_c}) and (\ref{PEc}) are 
asymptotically equivalent.

\section*{Appendix B: Proof of Theorem \ref{theorem2}}

{\it Proof}.  Assume that there exists a value $c>0$ such that $E[\tilde{\theta}_{}(c)]=\theta$. When $p=0.5$, by definition we have
\begin{align*}
{E}[\tilde{\theta}_{}(c)]
=\sum_{x=0}^n \left(\frac{n+2c}{x+c}\right) {n\choose x} p^x (1-p)^{n-x} = {1\over 2^n} h(c),
\end{align*}
where
\begin{align*}
h(c)=\sum_{x=0}^n \left(\frac{n+2c}{x+c}\right) {n\choose x}.
\end{align*}
Hence to show that the estimator is unbiased for $\theta=p^{-1}=2$, it is equivalent to show that there exists a value $c>0$ such that $h(c)=2^{n+1}$.

The first derivative of $h(c)$ is
\begin{align}
h'(c)
=\sum_{x=0}^n \frac{2x-n}{(x+c)^2} {n\choose x}.
\label{6}
\end{align}
When $n$ is an even number, by noting that ${n\choose x}={n\choose n-x}$, we can rewrite the first derivative as
\begin{align*}
h'(c) &= \sum_{x=0}^{n/2-1}\left[ \frac{2x-n}{(x+c)^2} {n\choose x}  +  \frac{2(n-x)-n}{(n-x+c)^2} {n\choose n-x} \right] \\
&= \sum_{x=0}^{n/2-1}\left[ \frac{2x-n}{(x+c)^2} {n\choose x}  +  \frac{n-2x}{(n-x+c)^2} {n\choose x}
 \right],
\end{align*}
where the term with $x=n/2$ is zero and so is excluded. Note also that, for any $x=0,\dots,n/2-1$, we have $n-x>x$ and further
\begin{align*}
\frac{2x-n}{(x+c)^2} {n\choose x}  +  \frac{n-2x}{(n-x+c)^2} {n\choose x} <\frac{2x-n}{(x+c)^2} {n\choose x}  +  \frac{n-2x}{(x+c)^2} {n\choose x} =0.
\end{align*}
This shows that $h'(c) < 0$ for any $c>0$.
When $n$ is an odd number, we can write the first derivative of $h(c)$ as
\begin{align*}
h'(c) = \sum_{x=0}^{(n-1)/2}\left[ \frac{2x-n}{(x+c)^2} {n\choose x}  +  \frac{2(n-x)-n}{(n-x+c)^2} {n\choose n-x} \right].
\end{align*}
And similarly, we can show that  $h'(c)<0$ for any $c>0$. Combining the above results, $h(c)$ is a strictly decreasing function of $c$ on $(0,\infty)$.

In addition, for any finite $n$ we note that
\begin{align*}
\lim_{c \to \infty} h(c)= \sum_{x=0}^n  \lim_{c \to \infty} \left(\frac{n+2c}{x+c}\right) {n\choose x} = 2\sum_{x=0}^n  {n\choose x} = 2^{n+1}.
\end{align*}
This shows that there does not exist a finite value of $c>0$ such that $h(c)=2^{n+1}$, and so Theorem 2 holds.

\section*{Appendix C: Proof of Theorem \ref{theorem3}}
{\it Proof}.
To prove (i), we note that $(n+c)/(x+c)$ is a rational function of $c$ and so is always continuous on the domain of $(0,\infty)$. Now since $n$ is also finite, $g(c)$ is a continuous function of $c$ on $(0, \infty)$.
Also for the limit of $g(c)$,
\begin{align*}
\lim_{c \to 0} g(c)
&=\lim_{c \to 0} \left[\sum_{x=0}^n \left(\frac{n+c}{x+c}\right) {n\choose x} p^x (1-p)^{n-x}\right]\\
&=\left(\lim_{c \to 0} \frac{n+c}{c}\right) (1-p)^{n} +
\sum_{x=1}^n\left[ \left(\lim_{c \to 0} \frac{n+c}{x+c}\right) {n\choose x} p^x (1-p)^{n-x}\right]\\
&=\infty,
\\
\lim_{c \to \infty} g(c)
&=\lim_{c \to \infty} \left[\sum_{x=0}^n \left(\frac{n+c}{x+c}\right) {n\choose x} p^x (1-p)^{n-x}\right]\\
&=
\sum_{x=0}^n\left[ \left(\lim_{c \to \infty} \frac{n+c}{x+c}\right) {n\choose x} p^x (1-p)^{n-x}\right]\\
&=1.
\end{align*}

To prove (ii), we verify that the first derivative of $g(c)$
\begin{align*}
g'(c)=\sum_{x=0}^{n-1} \frac{x-n}{(x+c)^2} {n\choose x} p^x (1-p)^{n-x}<0.
\label{a6}
\end{align*}
Hence, $g(c)$ is a strictly decreasing functon of $c$ on $(0, \infty)$.

To proof (iii), we show that the second derivative of $g(c)$
\begin{align*}
g''(c)
=\sum_{x=0}^{n-1} \frac{2(n-x)}{(x+c)^3} {n\choose x} p^x (1-p)^{n-x} >0.
\end{align*}
As a consequence, $g(c)$ is a strictly convex function of $c$ on $(0,  \infty)$.

\section*{Appendix D: Proof of Theorem \ref{theorem4}}

{\it Proof}. When $n=1$, equation (\ref{a7}) becomes
\begin{eqnarray*}
\left(\frac{1+c}{c}\right)(1-p)+p=\frac{1}{p},
\end{eqnarray*}
from which we obtain $c_1=p$.

When $n=2$, it is necessary to solve
\begin{eqnarray*}
\left(\frac{2+c}{c}\right) (1-p)^{2} + 2 \left(\frac{2+c}{1+c}\right ) p (1-p)
+   p^2
=\frac{1}{p}.
\end{eqnarray*}
After factorizing this equation, we have
\begin{eqnarray*}
c^2+(1-2p)c-2p(1-p)=0.
\end{eqnarray*}
The solutions are
$
c=p- {0.5} \pm \sqrt{{0.5} - (p-{0.5})^2}.
$
To remain a positive value of the estimator, the value of $c$ is required to be positive, so
$c_2=p- 0.5 + \sqrt{{0.5} - (p-{0.5})^2}$.

To get the solution of $c$ when $n\geq3$, we apply the Taylor expansion of $1/(X+c)$ around $c=1$ and it yields that
\begin{equation}
\frac{1}{X+c}
=\frac{1}{X+1} -  \frac{c-1}{(X+1)^2} + O((c-1)^2).
\label{appendix.18}
\end{equation}
By (\ref{g_c}) and (\ref{appendix.18}), for any finite $n$ we have
\begin{eqnarray}\label{formula26}
g_{}(c) &=&   {E}\left(\frac{n+c}{X+1}\right) - {E}\left[ \frac{(n+c)(c-1)}{(X+1)^2} \right] + O((c-1)^2) \nonumber \\
&=& {E}\left(\frac{n+c}{X+1}\right) - {E}\left[ {(n+1)(c-1) \over (X+1)^2} + {(c-1)^2 \over (X+1)^2} \right] + O((c-1)^2)\nonumber \\
&=& {E}\left(\frac{n+c}{X+1}\right) - {E}\left[ {(n+1)(c-1) \over (X+1)^2}  \right]  + O((c-1)^2).
\end{eqnarray}

Let $D_1=E[1/(X+1)]$ and $D_2 = E[1/(X+1)(X+2)]$. For $D_1$, we have
\begin{eqnarray}\label{D1}
D_1 &=& \frac{1}{n+1} \sum_{x=0}^n \frac{n+1}{x+1}{n \choose x}p^{x}(1-p)^{n-x} \nonumber \\
&=& \frac{1}{p(n+1)} \sum_{x=0}^n \frac{(n+1)!}{(x+1)!(n-x)!}p^{x+1}(1-p)^{n+1-(x+1)} \nonumber \\
&=& \frac{1}{p(n+1)} \sum_{s=1}^{n+1} {{n+1}\choose s}p^{s}(1-p)^{n+1-s} \nonumber \\
&=&{1\over p(n+1)}[1-(1-p)^{n+1}],
\end{eqnarray}
where $s=x+1$.
And for $D_2$, we have
\begin{eqnarray}\label{D2}
D_2
&=& \frac{1}{(n+1)(n+2)} \sum_{x=0}^n \frac{(n+1)(n+2)}{(x+1)(x+2)}{n \choose x}p^{x}(1-p)^{n-x}  \nonumber \\
&=& \frac{1}{p^2(n+1)(n+2)} \sum_{x=0}^n \frac{(n+2)!}{(x+2)!(n-x)!}p^{x+2}(1-p)^{n+2-(x+2)} \nonumber \\
&=&{1\over p^2(n+1)(n+2)}[1-(1-p)^{n+2}-(n+2)p(1-p)^{n+1}].
\end{eqnarray}
Now with $D_1$ and $D_2$, to derive the solution of $c$, we take the approximation $E[1/(X+1)^2] \approx (1+D_1)D_2$ and also ignore the remainder term $O((c-1)^2)$  in (\ref{formula26}). Then consequently, we have the approximate equation as $(n+c)D_1 - (n+1)(c-1)(1+D_1)D_2 \approx 1/p$, which yields
the approximate solution of $c$ as
\begin{eqnarray*}
c_n \approx   1 - {1/p-(n+1)D_1 \over  (n+1)(1+D_1)D_2-D_1}
= 1 - {p^{-1}(1-p)^{n+1} \over  (n+1)(1+D_1)D_2-D_1}.
\end{eqnarray*}

\section*{Appendix E: Proof of Theorem \ref{theorem5}}

{\it Proof}. For $c_n$ in formula (\ref{theorem3_formula}) with $p \in (0,1)$, by (\ref{D1}) and (\ref{D2}) we have
\begin{eqnarray*}
\frac{(n+1)(1+D_1)D_2-D_1}{(n+2)^{-1}}
&=&(1+D_1)\left[\frac{1}{p^2} - \frac{(1-p)^{n+2}}{p^2} - \frac{(n+2)(1-p)^{n+1}}{p} \right] \\
&&- {(n+2)\over p(n+1)}[1-(1-p)^{n+1}].
\end{eqnarray*}
Noting also that $\lim_{n\to \infty} D_1 = 0$, it leads to 
\begin{eqnarray*}
 (n+1)(1+D_1)D_2-D_1 = O\left(\frac{1}{n+2}  \right).
\end{eqnarray*}
Moreover, we have 
\begin{eqnarray*}
c_n \approx  1 - {p^{-1}(1-p)^{n+1} \over  (n+1)(1+D_1)D_2-D_1} = 1+ \frac{O[(1-p)^{n+1}]}{O[(n+2)^{-1}]} = 1 + o(1) ~~~~ {\rm as}~n\to \infty.
\end{eqnarray*}
Recall that the plug-in estimator $\tilde p_{\text {plug}}(\alpha_n)={\rm min}({\rm max}(\hat p_{\text {\tiny {MLE}}}, \alpha_n), 1-\alpha_n)$ with $\alpha_n =  1/(2+{\rm ln}(n))$ is bounded in $(0,1)$, then we have 
$\tilde c_n = c_n(\tilde p_{\rm plug}(\alpha_n )) = 1+ o_p(1)$. 
Finally, by the similar argument as in Theorem \ref{theorem1}, 
$\hat\theta_{}(\tilde c_n)$ is a consistent estimator of $\theta$. 

\newpage
\section*{Appendix F: Additional simulation results}


\begin{figure}[!hbt]
\centering
\includegraphics[width=14cm]{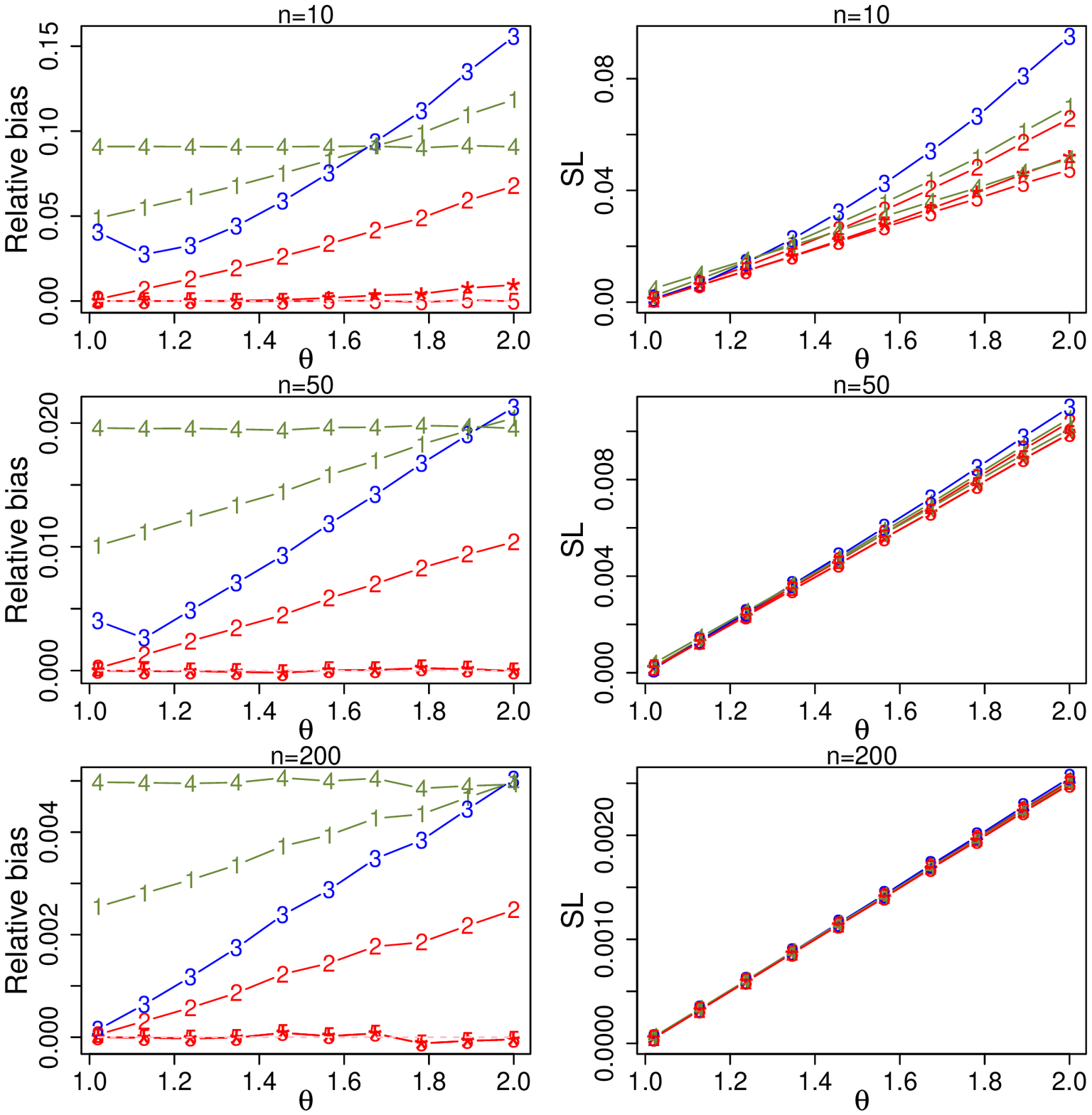}
 \caption{The relative biases and the Stein losses of the six estimators with $n=10$, 50 or 200 and $\theta \in [1.02,2]$, where ``\textasteriskcentered" represents the simulation results of our new estimator, ``1" represents the Walter estimator,
``2" represents the Pettigrew estimator, ``3" represents the piecewise Walter estimator, ``4" represents the Carter estimator, and ``5" represents the Fattorini estimator.}
\label{Additional-Figure-1}
\end{figure}

\end{spacing}
\end{document}